\documentclass[twocolumn,aps,prl]{revtex4}
\usepackage{graphicx}
\usepackage{amsmath}
\usepackage{amssymb}
\usepackage{wasysym}
\usepackage{color}

\newcommand{\SCA}{$\theta_{SCA}$}
\newcommand{\Rey}{${\cal R}e$}
\newcommand{\We}{${\cal W}e$}
\newcommand{\Rf}{{${\cal R}$$_f$}}
\newcommand{\unit}[1]{\ensuremath{\, \mathrm{#1}}}

\newlength{\textlarg}

\begin{document}

%=============== TITLE ================

\title{Drop impact upon micro- and nanostructured superhydrophobic surfaces}

\author{Peichun Tsai$^{\ddag}$, Sergio Pacheco$^{\S}$, Christophe Pirat$^{\ddag,\dag}$, Leon Lefferts$^{\S}$, and Detlef Lohse$^{\ddag}$. }

\affiliation {$^{\ddag}$Physics of Fluids Group; $^{\S}$Catalytic Processes and Materials Group, 
Faculty of Science and Technology, University of Twente, 7500AE
Enschede, The Netherlands.\\
$^{\dag}$Present address: Laboratoire de Physique de la Mati\`ere Condens\'ee et Nanostructures,
Universit\'e de Lyon; Univ. Lyon I, CNRS, UMR 5586, 69622 Villeurbanne, France}
\date{\today}

\begin{abstract}
We experimentally investigate drop impact dynamics onto different superhydrophobic surfaces, consisting of regular polymeric micropatterns and rough carbon nanofibers, with similar static contact angles. The main control parameters are the Weber number \We~and the roughness of the surface.  At small \We, i.e. small impact velocity, the impact evolutions are similar for both types of substrates, exhibiting Fakir state, complete bouncing, partial rebouncing, trapping of an air bubble, jetting, and sticky vibrating water balls. At large \We, splashing impacts emerge forming several satellite droplets, which are more pronounced for the multiscale rough carbon nanofiber jungles. The results imply that the multiscale surface roughness at nanoscale plays a minor role in the impact events for small \We~$\apprle 120$ but an important one for large \We~$\apprge 120$. Finally, we find the effect of ambient air pressure to be negligible in the explored parameter regime \We~$\apprle 150$.
\end{abstract}

\maketitle

% ==================
\section{Introduction}
% ==================

Drop impact on a solid surface is ubiquitous and crucial in a variety of industrial processes.  The interplay of several experimental conditions often makes the impact dynamics surprising and too complex to elucidate~\cite{Yarin_review}.  The control parameters include the impact velocity, liquid density, viscosity, surface tension, and the wettability and roughness of the substrates.  In the past decade, the investigations have actively focused on the impacts upon superhydrophobic substrates~\cite{Quere_review,DQuere_EPL2000,DQuere_Nature_2002,Quere_PRE_2004,bartolo_jets_PRL2006,DQuere_bouncing_transition_EPL2006,JungYongChae_la2008, JTimonen_EPL2008} due to their emergence and opportunities in a wide range of applications: for instance, self-cleaning and anti-coating for lab-on-chip devices, ink-jet printing, spraying techniques, and coating processes. The most commonly studied substrates are silane-coated microstructures and hydrophobic microtextures. These samples, possessing a combination of chemical hydrophobicity and physical ordering roughness, display superhydrophobicity--with large static contact angle above $150^\circ$ and small contact angle hysteresis within $5^\circ$.  

In this study, we investigate the dynamics of drop impact not only on superhydrophobic microstructures of controlled roughness but also on multiscale rough surfaces of carbon nanofiber jungles (CNFJs).  The aim is to compare the impact behaviors upon both types of superhydrophobic surfaces and to examine the effect of the details of the roughness.  Moreover, complementary to recent studies of drop impact using superhydrophobic surfaces, our work examines the pressure of the surrounding air, as an additional control parameter, which recently has been shown to affect drop impact upon dry wetting surfaces~\cite{xu:splashing_PRL2005, xu_PRE2007_1,xu_PRE2007_2}.

%[about PDMS]
Chemical and/or geometrical surface modification affects the solid-liquid interactions.  This offers a convenient control in surface physics. For example, physical roughness increases (decreases) the contact angle for hydrophobic (hydrophilic) substrates. Recently, superhydrophobic surfaces through controlled microstructures (see Fig.~\ref{sem_pics}c) have been developed, advanced, and investigated for their water-repelling~\cite{fractal_surfaces, DQuere_Nature_2002, Quere_nature_2003} and hydrodynamic slip applications~\cite{ECharlaix_2007, ELauga_noslip_BC_2007}. A water droplet when deposited on a micropatterned hydrophobic surface can exhibit a metastable state of heterogeneously wetting with air trapped between the liquid and surface, forming a ``Fakir'' drop~\cite{DQuere_NatureMat_2002,DQuere_2005}.  A spontaneous transition can occur from ``Fakir'' to homogeneously wetting ``Wenzel" state with a smaller contact angle, even in the case of zero impact velocity~\cite{Sbragaglia_PRL2007,HStone_nature_2007}.  At the breakdown of superhydrophobicity, intriguing filling dynamics of water infiltration was found to possess multiple timescales and to tune the shape of the wetted areas by the geometric dimensions of the micropillars~\cite{Pirat_EPL2008}.  In drop impact experiment, the threshold of the impact velocity $V_c$ marking the transition from Fakir to Wenzel state is found to depend on the geometric dimensions of the hydrophobic microstructures~\cite{DQuere_bouncing_transition_EPL2006, JTimonen_EPL2008}.  So far, geometric arrangements and the impact velocity have been the main control parameters for the drop impact onto superhydrophobic surfaces~\cite{DQuere_EPL2000, bartolo_jets_PRL2006}.  The change in surrounding air pressure has not yet been explored in this context. We will also study this effect here.

%[ About CNFJs]
The analyzed superhydrophobic substrates consist of carbon nanofiber jungles.  Carbon filaments are formed catalytically in metallic catalysts, particularly in Ni, Fe and Co based catalysts, used for the conversion of carbon containing gases, e.g. in steam reforming of hydrocarbons and Fischer-Tropsch synthesis~\cite{CNF_Ref2}. In these cases, the carbon filament formation was detrimental for operation as they plugged reactors and deactivated catalysts. 
Fiber type carbon nanomaterials can be classified into three types, namely, Carbon Nanofiber (CNF), Carbon Nanotube (CNT) and Single Walled Nanotube (SWNT). In CNF the graphitic planes are oriented at an angle to the central axis, thus exposing graphite edge planes. In CNT the graphitic planes run parallel to the central axis, in this state only basal planes are exposed. CNT is also referred to as Multi-Walled Carbon nanotubes (MWNT) or parallel carbon nanofibers. If the fiber consists of only one graphene sheet that is oriented in the direction parallel to the fiber axis, it is called a single-walled carbon nanotube.
We refer the reader to Ref.~\cite{CNFs,CNF_synthesis_application,CNF_review} for the history, studies, and reviews of catalytically generated CNFs.

The diameters of CNFs, in general, range from a few to hundred nanometers and their lengths vary from micrometers to millimeters~\cite{CNFs}.  The novel physical and chemical properties of CNFs include high surface areas, large elasticity, and low electrical resistivity and hence make CNFs good candidates for catalysts, catalyst supports, and selective adsorption sites~\cite{CNF_review}.  Particularly, the highly porous layers of entangled nanofibers (CNF jungles) are suitable as catalyst supports for liquid phase reaction, allowing fast mass transfer to catalytic active particles deposited on the CNFs~\cite{CNF_Ref30,CNF_Ref50}. In this work entangled layers of CNFs are grown directly on iron and stainless steel metal foils, and we experimentally study the dynamical effect of water drop impinging upon these CNFs.

% =====================
\section{Experimental Section}
% =====================

\subsection{Superhydrophobic samples}

\begin{figure}
\begin{center}
\includegraphics[width=3in]{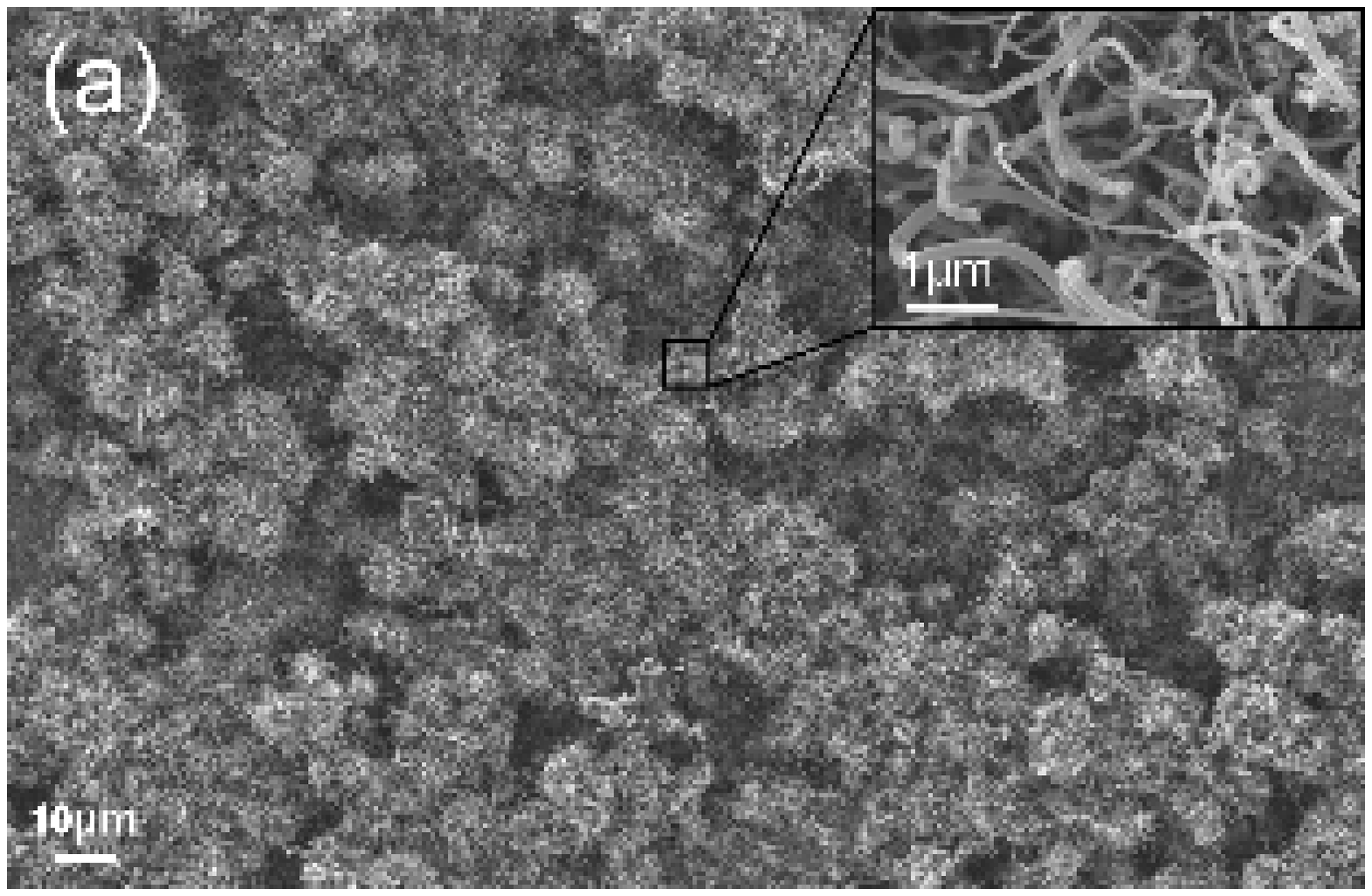}
\includegraphics[width=3in]{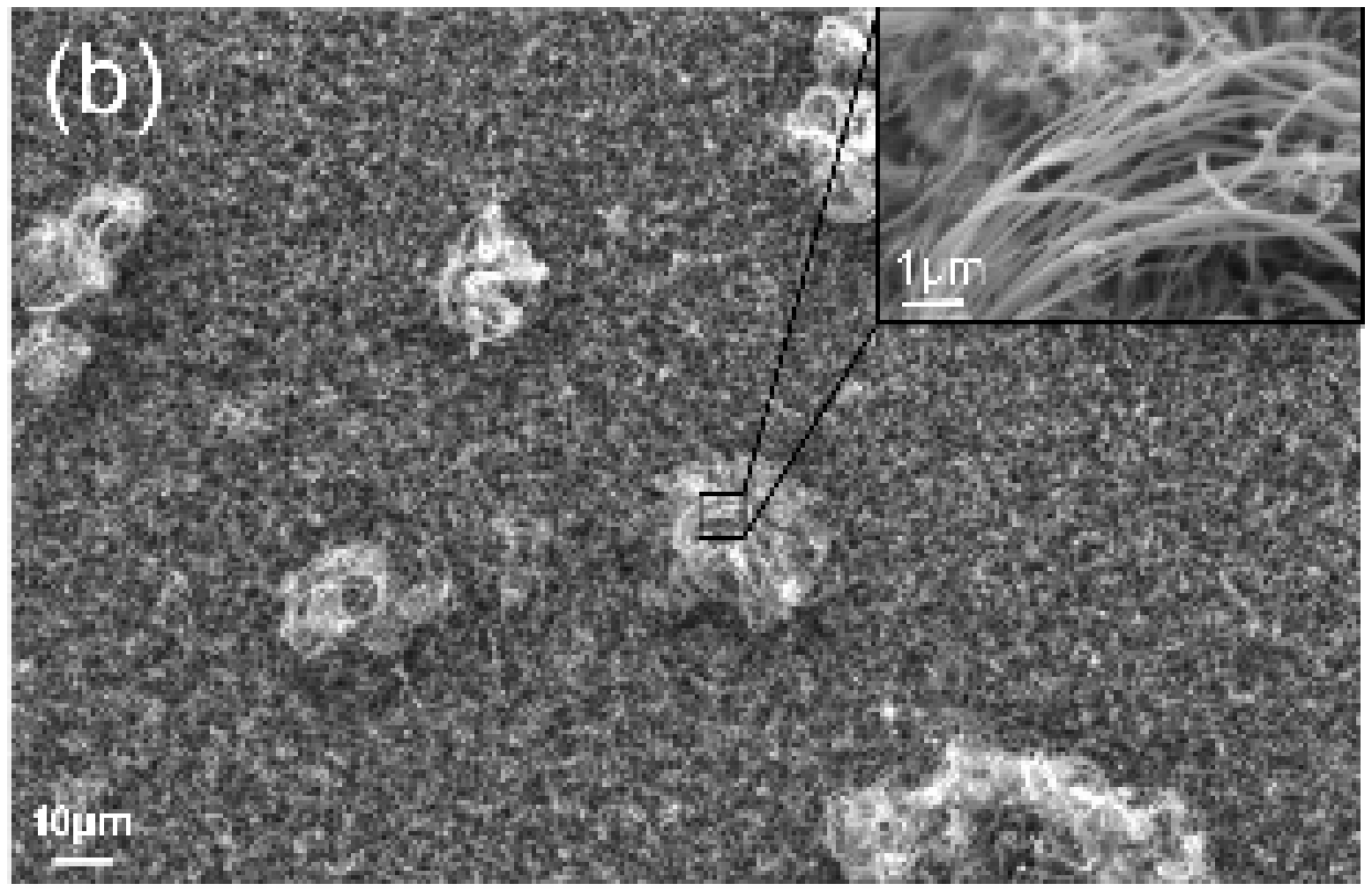}
\includegraphics[width=3in]{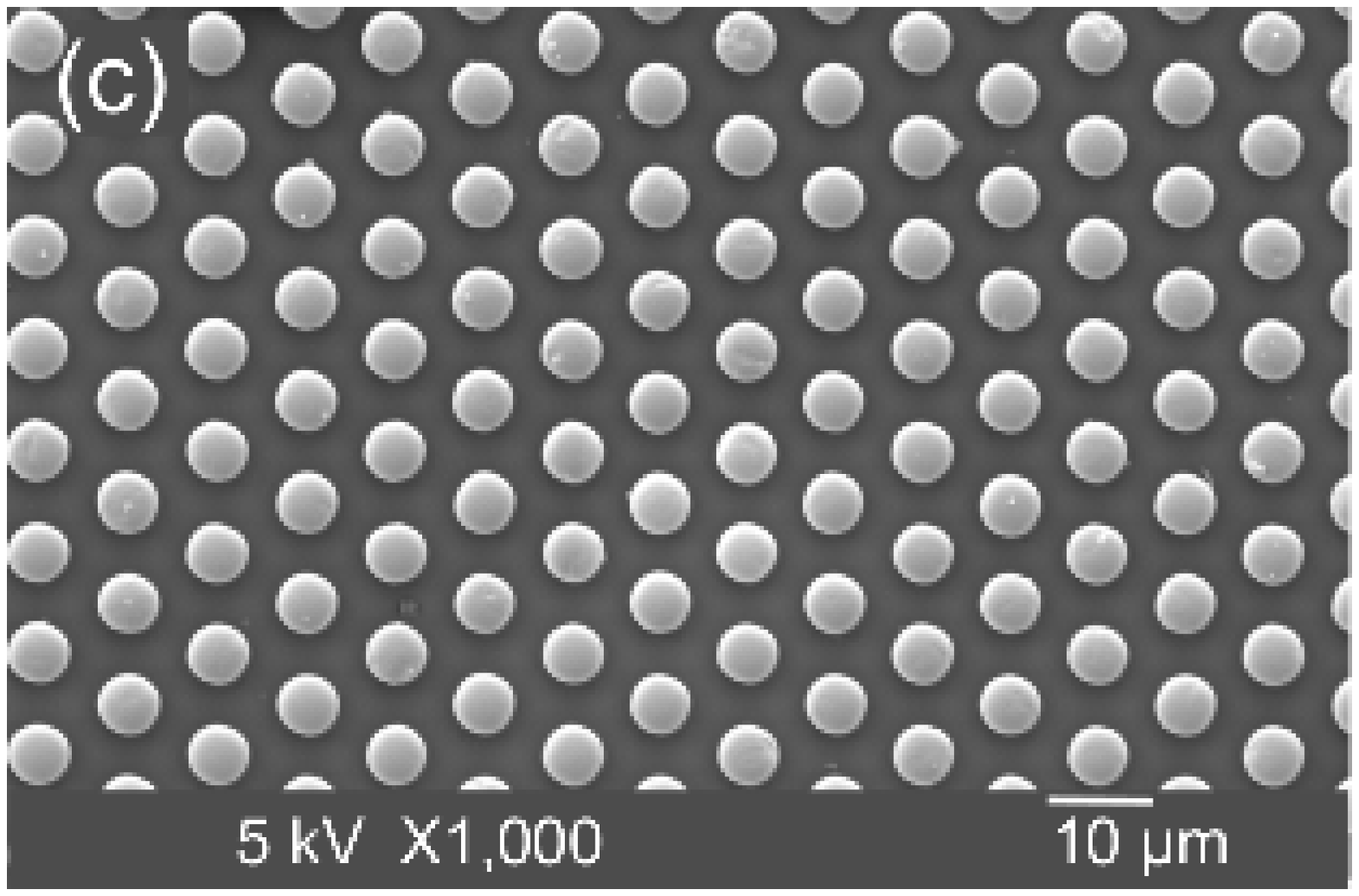}
\caption{\label{sem_pics}(a) and (b): SEM pictures of two representative used samples of carbon nanofiber jungles, showing uncontrolled, multiple-scale roughness. The porous fiber-bundles span a few tens microns and the fine fibers are submicron thick. (c) A SEM top-view image of a representative microstructured substrate consisting of $5~\unit{\mu m}$ wide pillars, yielding a well-defined roughness by the geometric regularity~\cite{micro-molding}.
}% (a) Fe. and (b) stainless steel as a catalysts. 
\end{center}
\end{figure}

Our superhydrophobic substrates are catalytically synthesized CNFJs and micropatterned polymers. Fig.~\ref{sem_pics} shows representative scanning electron microscope (SEM) images of used samples. As revealed by Fig.~\ref{sem_pics} (a) and (b), catalytically generated CNFJs are composed of entangled and bundled carbon nano-filaments forming uncontrolled, microscopic roughness. In contrast, Fig.~\ref{sem_pics} (c) shows the top-view of an ordered, microstructured substrate comprising micro-pillars periodically placed in a structured lattice.

% \subsection{Synthesis of Carbon Nanofiber Jungles}
%[ Procedures of making CNFJs ]
{\bf Synthesis of Carbon Nanofiber surfaces.} 
The carbon nanofibers (CNFs) were produced by catalytic vapor deposition (CVD) from carbon containing gases using a metallic catalyst~\cite{CNF_review}.   
% Materials
Two types of catalytic materials were employed: they are iron ($99.99\%$, Alfa Aesar) and stainless steel Type 304 (Fe:Cr:Ni 70:19:11\%, Alfa Aesar) foils of $0.1$~mm thick. Round samples of these metal foils ($10$~mm in diameter) were prepared by an electric discharge wire-cutting machine (Agiecut Challenge 2, GF AgieCharmilles). The foils were degreased ultrasonically in acetone and dried at room temperature before loaded into the CVD chamber. In addition, hydrogen and nitrogen ($99.999\%$ purity, Praxair), and ethylene (C$_2$H$_4$, $99.95\%$ purity, Praxair) were used for the CNF formation without further purification.

%Carbon Nanofibers formation
The CVD reactor consists of a vertical quartz reactor, with a porous quartz plate centrally placed to support the metal foils. The temperature was raised from room temperature to $600^{\circ}$C at a rate of $5$K/min. The samples were first pre-treated in a hydrogen/nitrogen mixture ($20\%$ H$_2$ with $80\%$ N$_2$) with a total flow of $100~\unit{ml/min}$ for 1 hr at $600^{\circ}$C. After the pre-treatment, ethylene was fed into the reactor (20\% with N$_2$/H$_2$, $20\%/60\%$) at $600^{\circ}$C for $2$ hrs (with stainless steel foil) or $3$ hrs (with iron foil). During the heating process at the high temperature, the decomposition of C$_2$H$_4$ led to the growth of carbon filaments upon the metallic catalysts. The concentration of hydrogen and the total flow were kept the same during pre-treatment and deposition. Finally the ethylene and hydrogen gas streams were shut off and the whole system was cooled down to room temperature under nitrogen at a rate of $10$ K/min.  

% \subsubsection{Preparation of the polymeric microstructures}
%[Preparation of the PDMS samples ]
{\bf Preparation of the polymeric microstructures.}
Micro-patterned substrates are composed of PDMS elastomer (Polydimethylsiloxance, RTV 615 rubber component A and curing agent B, GE Bayer Silicones).  The fabrication of precise and controllable microstructures was achieved via a micro-molding method~\cite{micro-molding}. This technique is to cast a polymeric film from solution on a molding wafer of desirable micropatterns.  The PDMS films were obtained by mixing the rubber component A with the curing agent B ($10 : 1~wt/wt$). The mixture was degassed and then poured onto the mold ($\sim 1$ mm) and cured in an oven for $3$ hrs at $85^\circ$C.  The casting molds consisted of diverse arrays of micro-patterns, which can produce periodically arranged, round or rectangular polymeric micro-pillars of height $h$, diameter or width $w$, and the interspacing $a$. With different lattice arrangements, $h$, $a$, and $w$, we controlled the roughness of the polymeric microstructures. We characterize the surface roughness \Rf~by the ratio of the total surface area to that projected on the horizontal plane  For the experiments presented here, $w = 5~\unit{\mu m}$, $h = 6,~10$, and $20~\unit{\mu m}$, and $a$ is varied between $1.27~\unit{\mu m}$ and $5~\unit{\mu m}$.  The corresponding \Rf~spans between $1.5$ and $9$.

{\bf Contact Angle Measurements.}
The contact angle measurements were performed with a milli-Q water droplet of $4~\unit{\mu l}$, matching the volume of impacting droplet of $\sim 1~\unit{mm}$ wide in radius, with the Laplace-Young fitting method (OCA 20, DataPhysics).  Catalytically generated CNFJs exhibit random topography of multiscale roughness, and thus we need to ensure the homogeneous wettability of such surfaces when performing drop impact experiments. The static contact angle $\theta_{SCA}$ of CNFJs reported here was measured at different locations on the same sample, both before and after the drop impact experiment. The scattering of these data is used to estimate the error in $\theta_{SCA}$.

\subsection{Drop impact experiments}

The general experimental procedure consists in releasing an impinging milli-Q water droplet from a fine needle ($0.1$ mm inner diameter) with a syringe pump (PHD 2000 Infusion, Harvard Apparatus) at different heights to vary the impact speed upon the solid surface. The balance between the surface tension and droplet gravitational force sets the droplet size, which is $\approx 1$ mm in radius within $5\%$ deviation.  The whole experimental setup was enclosed in a chamber, connected with a vacuum pump so as to control the pressure of the surrounding air. The dynamics of drop impact was recorded by a high-speed camera (Fastcam SA1, Photron) with a recording rate ranging from $1500$ to $30~000$ fps (frames per second).  The impact velocity and the droplet size were determined from the captured images.

%Control parameters
Several control parameters influence the impact dynamics, for example, droplet size, liquid viscosity $\mu$, impact velocity $V_i$.  We describe these effects in terms of dimensionless numbers: the Weber number ${\cal W}e$, the ratio of kinetic energy to surface energy, characterizing the deformability of the droplet; the Reynolds number \Rey,~the ratio of inertia to viscosity effect: 
\begin{equation}
{\cal W}e = \frac{\rho R V_i^2}{\sigma},\qquad {\cal R}e = \frac{\rho R V_i}{\mu}.
\end{equation}
Here $R$ is the radius of the liquid drop, $V_i$ is the impact velocity, $\rho$ is the liquid density, $\sigma$ is the surface tension, and $\mu$ is the liquid viscosity.  In this paper, we did not simultaneously change \We~and \Rey~by using different liquids; instead, these parameters are only different ways to non-dimensionalize the velocity $V_i$ of a milli-Q water droplet.   The Ohnesorge number, comparing viscous and capillary forces, {\bf Oh} = $\sqrt{{\cal W}e}/{\cal R}e = \mu/\sqrt{\rho R \sigma}$ is small $\sim O(10^{-3})$ in our experiments.
In addition, we control and measure the pressure of the surrounding air, noted as $P_{air}$. For the data presented here we specify the cases under reduced air pressure, otherwise $P_{air} = 101.3~\unit{kPa}$ at the standard ambient pressure. 

% =======================
\section{Results and Discussion}
% =======================

%\subsection{Drop impact upon carbon nanofiber forests}
%\subsubsection{The effect of Weber number}
\begin{figure*}
\begin{center}
\includegraphics[width=6.6in]{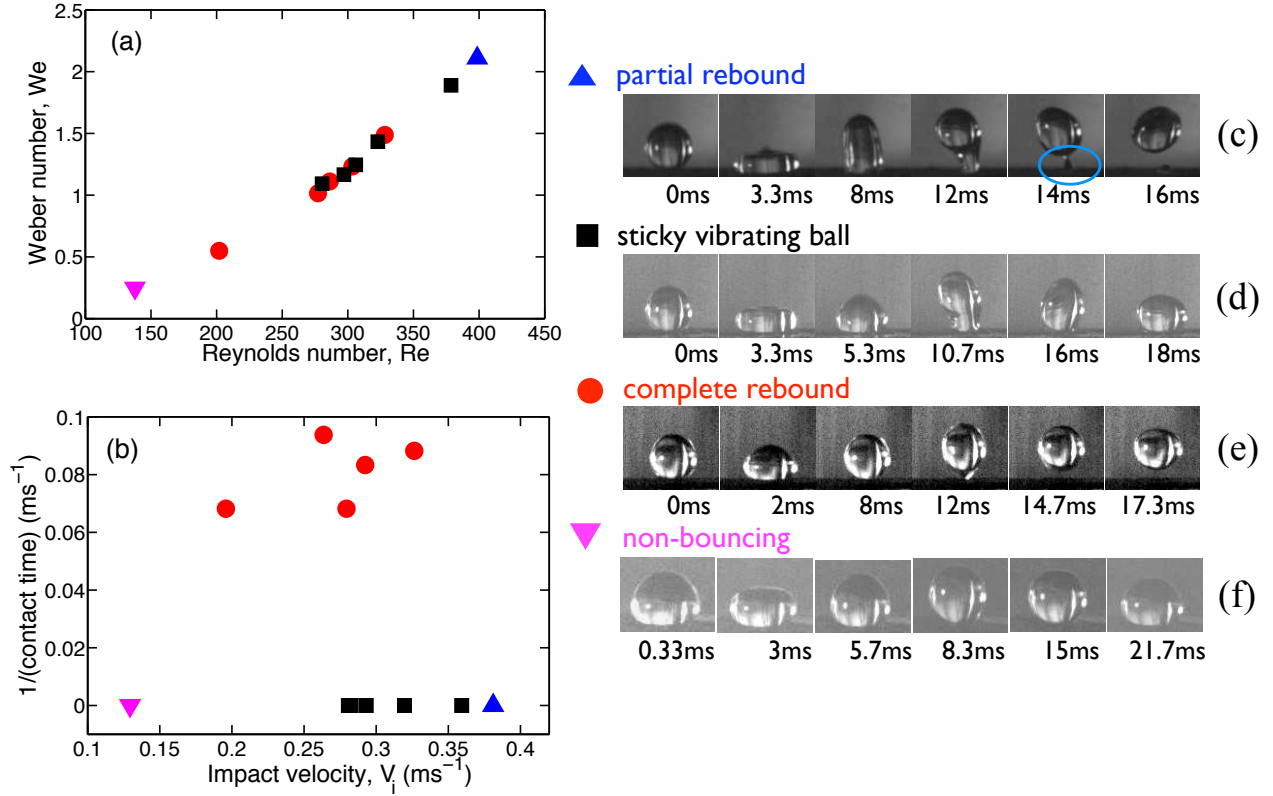}
\caption{\label{CNFs_SPB15_small_We}(a) Fraction of the phase space of the impact dynamics upon a carbon nanofiber substrate (Fig.~\ref{sem_pics}a) with a static contact angle of $(155 \pm 3)^{\circ}$ % (with the Laplace-Young fitting)
 at small \We~numbers, with the time evolutions of the impact events shown in (c)-(f). (b) shows the corresponding impact velocity $V_i$ vs. the inverse contact time.}
\end{center}
\end{figure*}

In this section, we compare impact events on CNFs and on microstructured surfaces with similar wettability. 
Fig.~\ref{CNFs_SPB15_small_We} shows  a small fraction of the phase space and the time evolution of the impact dynamics for water drops impacting upon CNFJs of $
\theta_{SCA} = 155 \pm 3^{\circ}$.  In this small \We-number regime ($\apprle 2.5$), the impact events include non-bouncing Fakir droplet, complete rebound, sticky wetting ball, and partial rebound, generally in this sequence as \We~is gradually increased.  When a water droplet is gently deposited, due to small kinetic energy the droplet maintains in a Cassie-wetting ``Fakir'' state with air trapped underneath the drop and thus a high contact angle during the whole impacting process, as shown in Fig.~\ref{CNFs_SPB15_small_We}f. In this heterogeneous state, air is trapped underneath the droplet and between the porous fibers. 
The non-wetting nature of the sample limits the spreading of the drop and sometimes even leads to a complete rebound (Fig.~\ref{CNFs_SPB15_small_We}e). 
At high $V_i \apprge 0.28 \unit{ms}^{-1}$ a wetting transition can occur as the kinetic energy overcomes the surface energy, associated with the liquid surface tension, and the droplet turns to the completely wetting Wenzel state, in which it is pinned on the surface (Figs.~\ref{CNFs_SPB15_small_We}c and d).

The average contact time before the droplet bounces off the surface is $12.5~ms$ (based on the marked data ($\bullet$) in Fig.~\ref{CNFs_SPB15_small_We}b).  This average value is consistent with the value $14~\unit{ms}$ found by a recent study with a multiwalled carbon nanotube array with a static contact angle of $163^{\circ}$~\cite{zwang_APL2007}.  
The coexistence of different dynamical behaviors for $0.28~\unit{ms^{-1}} \apprle V_i \apprle 0.35~\unit{ms}^{-1}$ illustrates the complexity of the impact process, showing the interplay of several parameters and indicating that other factors can affect the phase space. For instance, from our results we note that the complete rebound regime happens at higher $V_i$ between $0.5~\unit{ms}^{-1}$ and $1.1~\unit{ms}^{-1}$ for a more hydrophobic CNF substrate with a larger \SCA $=(163 \pm 4)^{\circ}$.

\begin{figure}
\begin{center}
\includegraphics[width=3.6in]{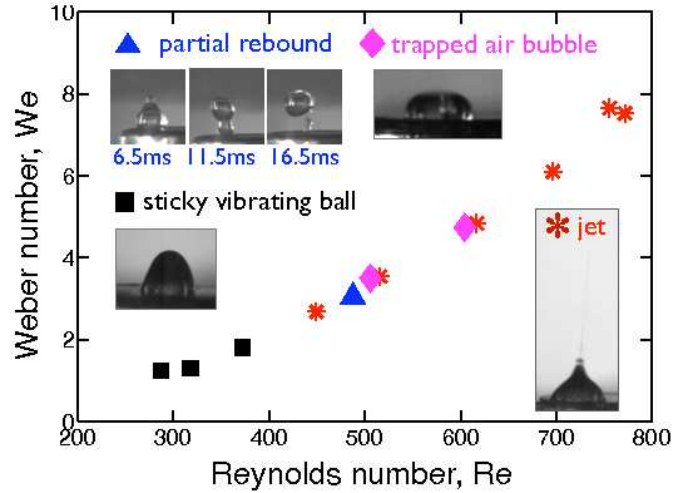}
\caption{\label{CNFs_moderate_We}(a) Impact dynamics upon a substrate of CNFJs,
 (shown in Fig.~\ref{sem_pics}b) with a static contact angle of $(152\pm3)^{\circ}$. The insets show the snapshots of the impact events: sticky ball at the Wenzel state ($\blacksquare$), partial rebound ($\blacktriangle$), trapping of an air bubble ($\blacklozenge$), and jetting ($\ast$). 
}
\end{center}
\end{figure}

\begin{figure*}
\begin{center}
\includegraphics[width=6in]{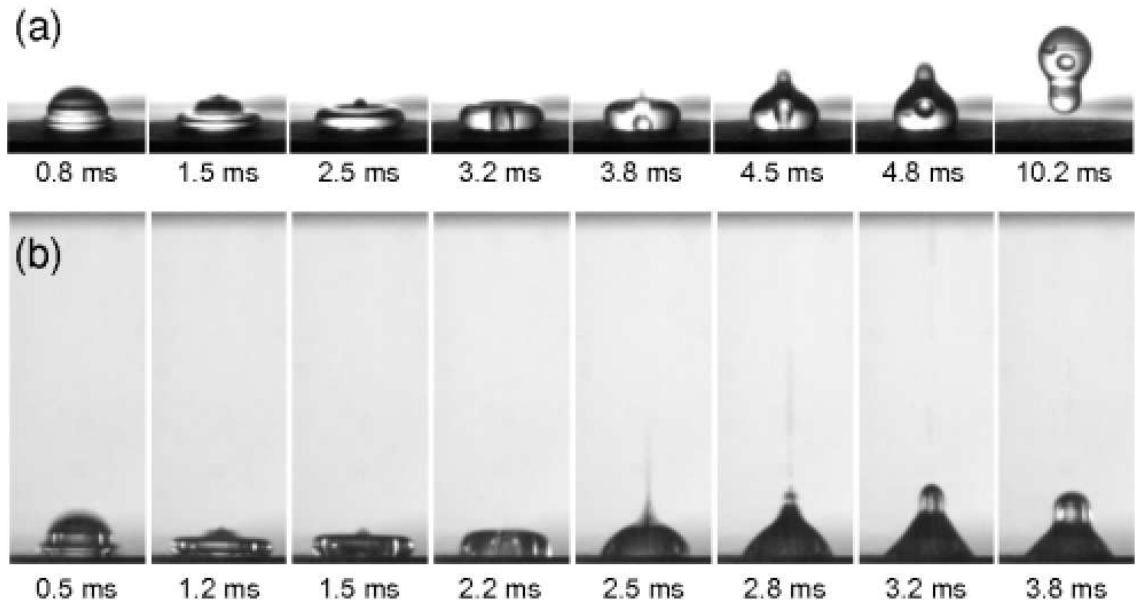}
\caption{\label{jets_and_bubbles}Time evolutions of the impact events of (a) trapping an air bubble at $\theta_{SCA} = (163 \pm 3)^{\circ}$, \We~$= 5.3$ and \Rey $ = 571$, and (b) jetting at $\theta_{SCA} = (152 \pm 3)^{\circ}$, \We~$= 7.7$ and \Rey~$ = 756$.}
%(a) bubble: 072208_exp7_CNF69; (b) 021508_exp1_CNF34.
\end{center}
\end{figure*}

Fig.~\ref{CNFs_moderate_We} shows the phase diagram for \We~between $1$ and $10$, corresponding to high \Rey, when a water droplet impacts upon a CNF surface of $\theta_{SCA} = (152 \pm 3)^{\circ}$ shown in Fig.~\ref{sem_pics}b.  In this \We-range, as discussed above, the impact scenario can display a complete bouncing or a pinning of contact line resulting in wetting.  Also, trapping an air bubble or jetting can happen due to the development of an air cavity.  Fig.~\ref{jets_and_bubbles} shows the detailed evolutions of (a) trapping an air bubble and (b) emitting a fast jet. For our \We-range between $2$ and $8$, on impact a surface capillary wave is excited and the droplet deforms like a pyramid around $1.5$ ms.  The oscillation of the surface deformation often makes a toroidal droplet producing a cylinder-like cavity; see the snapshot at $3.2~ms$ in Fig.~\ref{jets_and_bubbles}a and that at $2.2~ms$ in Fig.~\ref{jets_and_bubbles}b.  Then the following dynamics of the cavity restoration determines the subsequent behaviors: in (a) at a lower $V_i$ the surface wave restores the top of cavity sooner and closes it up producing a residual void; in (b) at a higher $V_i$ the fast collapse of the cavity produces a fast jet. Our data reveal that the shooting jets can be as fine as $30 \unit{\mu m}$ in radius and as fast as $9\times~V_i$.  Similar observations underlying the same mechanism have recently been investigated with ordered, superhydrophobic microstructures for the \We-number ranging between $0.6$ and $16$~\cite{bartolo_jets_PRL2006}. The analogy of the impact phenomena suggests that the formation and the collapse of the air cavity may be insensitive to the details of surface roughness.

\begin{figure*}
\begin{center}
\includegraphics[height=3.2in]{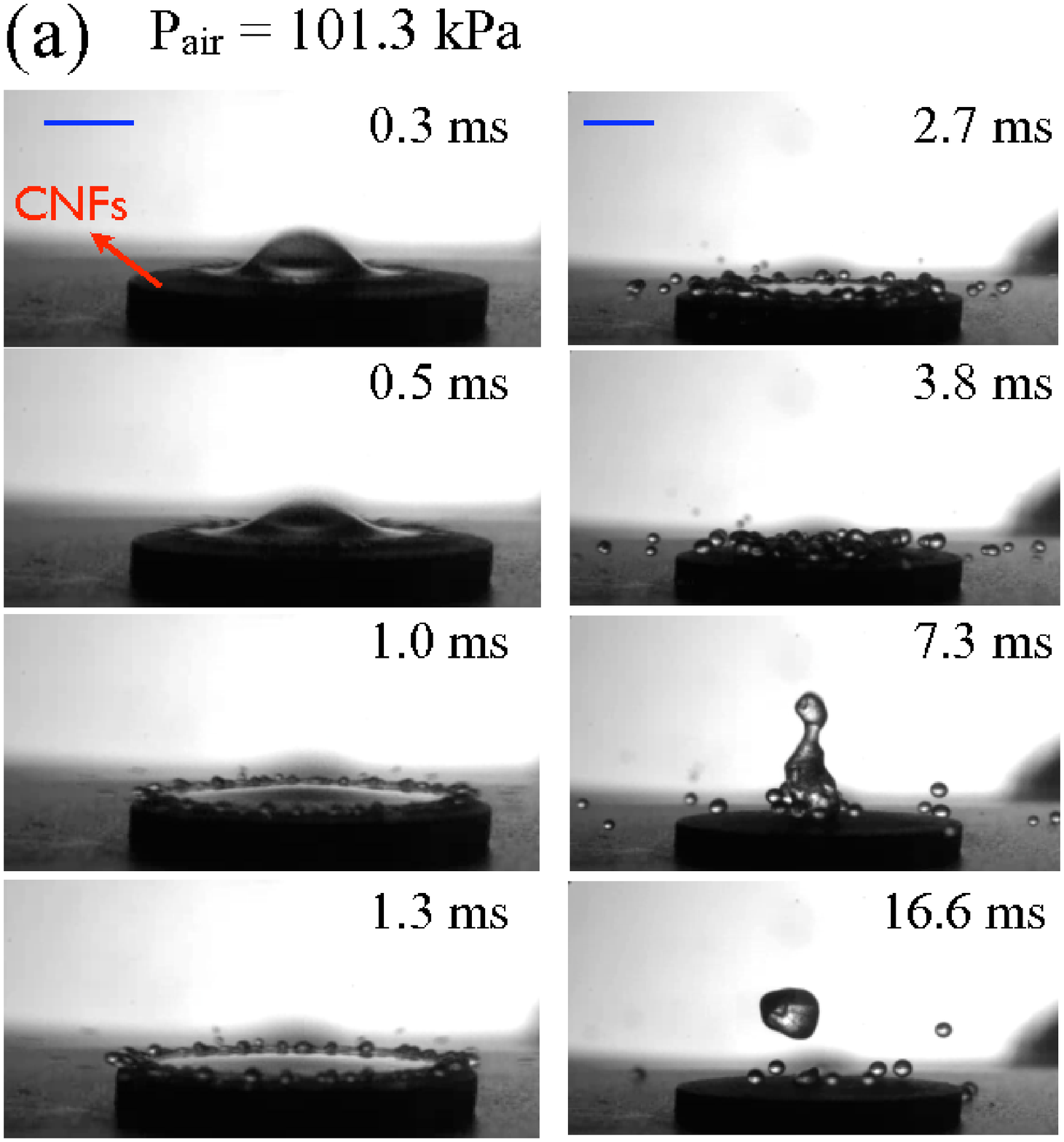}
\includegraphics[height=3.2in]{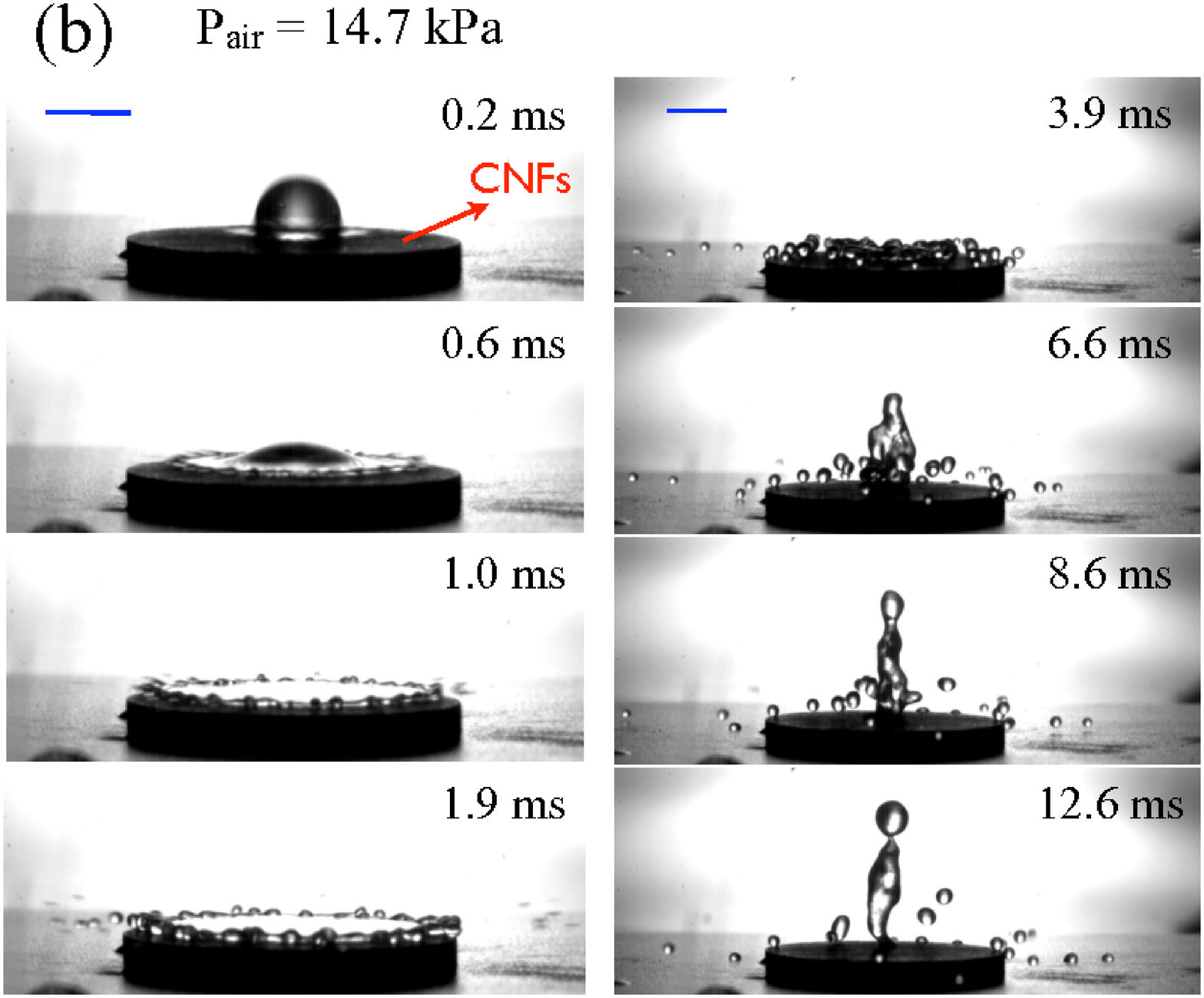}
\caption{\label{splashing_figs}Snapshots of the splashing evolutions on CNFs of $\theta_{SCA} = (163\pm3)^{\circ}$ at different air pressure (a) $P_{air} = 101.3~kPa$, \We~$ = 115.3$ and \Rey~$= 2784$, and (b) $P_{air}~=~14.7~kPa$, \We~$=~141.7$ and \Rey~$=~3060$. The inset bars indicate a length scale of $2$~\unit{mm}.}
% (a) 072308_exp10_SPB_69_Pred0mbar; (b) 072408_exp9_CNF_SPB69_Pred866mbar.
\end{center}
\end{figure*}

In the higher \We-number range (between $90$ and $140$) splashing impacts occur as shown by Fig.~\ref{splashing_figs} for CNFs. Here, the ``splashing'' phenomenon refers to the formation of many satellite droplets, which merge during the spreading and/or contracting stages of the water film from our observations.   In Fig.~\ref{splashing_figs}a under $1~atm$, at 1 ms tiny droplets radially emit while the water sheet still spreads outwards with wavy perturbations, resembling ``fingers'', at the edge of water film. 
Several droplets originated from the wavy perimeter develop between $1.8~ms$ and $5.8~ms$ while the main water sheet contracts, shaping into a partial rebound. 
A similar splashing phenomenon has lately been observed at \We~$\approx 160$ with a superhydrophobic surface consisting of regular micro-pillars of $h = 37 \unit{\mu m}$ in height, and $a = 3 \unit{\mu m}$ in the interspacing distance between pillars~\cite{DQuere_bouncing_transition_EPL2006}.  This Ref. also reports that the critical \We-number above which droplets eject is about $700$ for a flat hydrophobic solid and no fragmentation of droplets even at \We$~= 1000$ for a flat hydrophilic solid.  With these observations, the authors conjecture that the role of the film of air plays a crucial role because the non-wetting Fakir state promotes an air film for superhydrophobic solids. However, our investigations find the effect of air pressure on the splashing to be negligible in this parameter regime; Fig.~\ref{splashing_figs}b reveals the same kind of splashing at a low air pressure, $P_{air}=14.7~kPa$, with similar \We~and \Rey~to those in Fig.~\ref{splashing_figs}a.  Our experiments at different $P_{air}$ with similar \We~numbers resemble the same type of splashing. Hence, the data suggest that the mechanism of aforementioned splashing behaviors is due to the effect of surface roughness, and not due to the surrounding air. 

\begin{figure*}
\begin{center}
\includegraphics[width=6.5in]{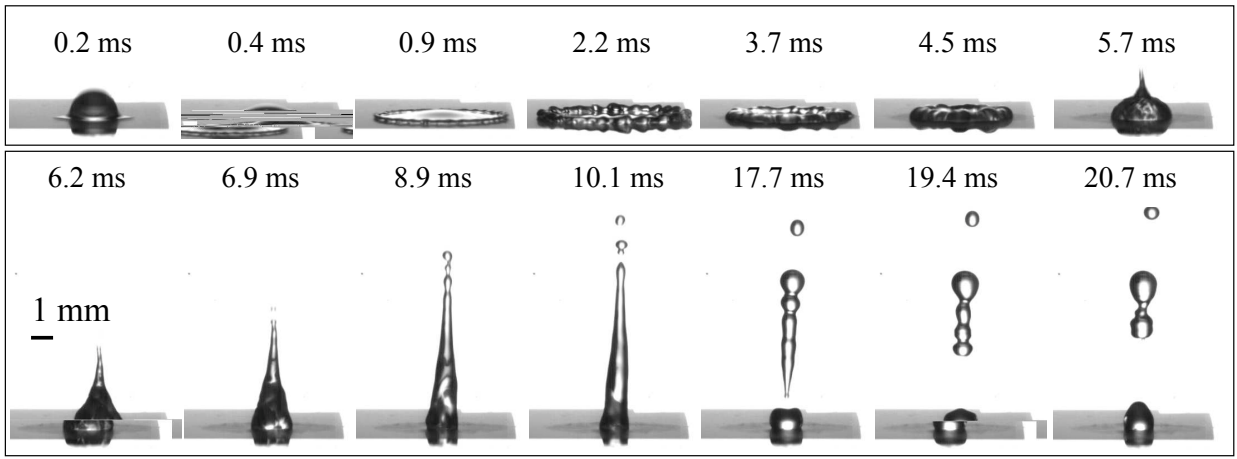}
\caption{\label{PDMS_We120}Time evolutions of impact dynamics upon an ordered, microstructured polymeric substrate of $\theta_{SCA}=(147 \pm 3)^{\circ}$ at $P_{air} = 101.3~\unit{kPa}$, \We~$= 120$ and \Rey~$= 2882$. This surface is comprised of round pillars of $a=1.27~\unit{\mu m}$, $w=5~\unit{\mu m}$ and $h=10~\unit{\mu m}$ arranged in a rectangular lattice and yields the roughness \Rf~$= 5$.}% 072408_exp6_partial rebound_PDMS2924.
\end{center}
\end{figure*}

\begin{figure}
\begin{center}
\includegraphics[width=3.6in]{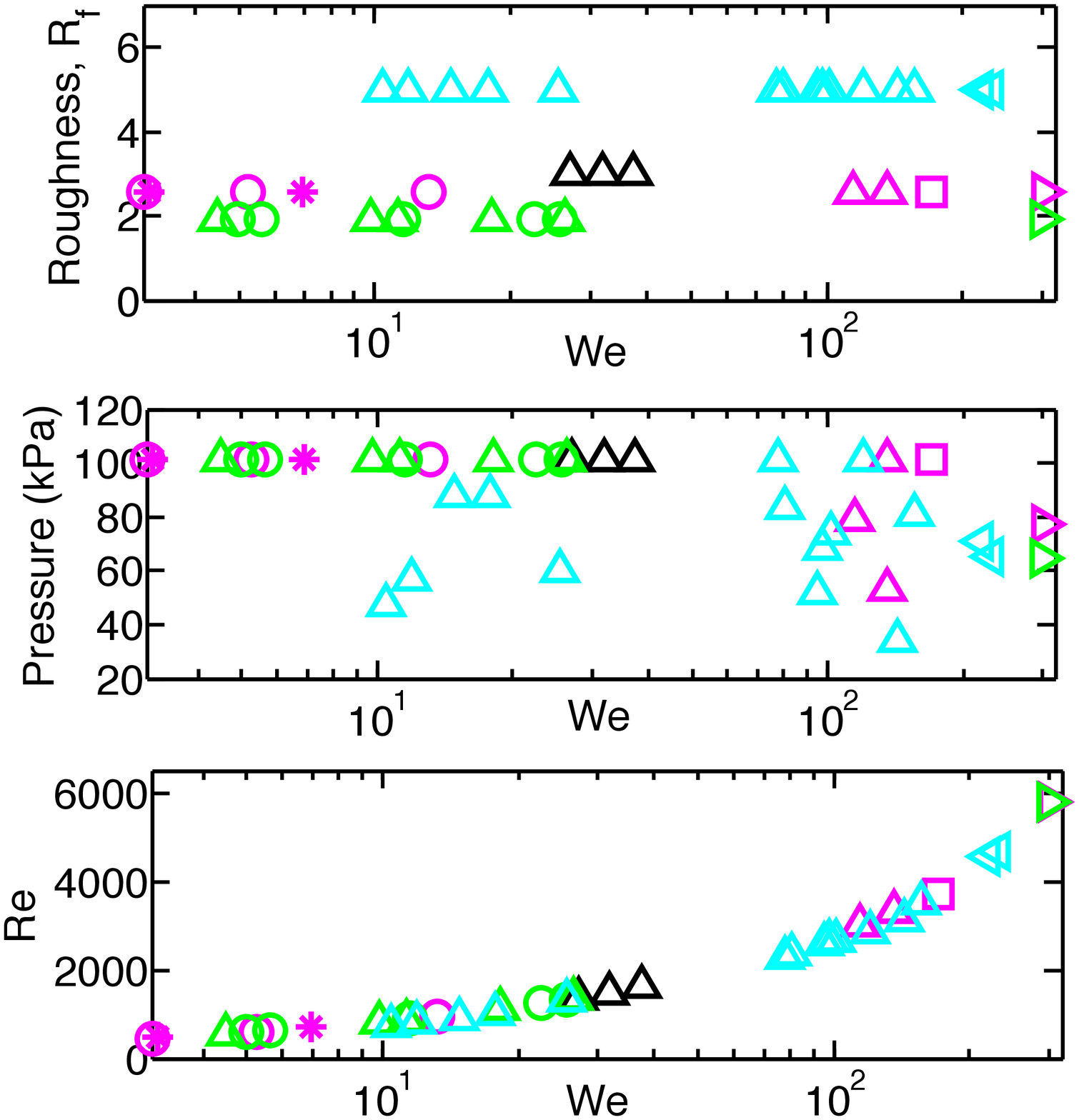}
\caption{\label{PDMS_phase_diagram}Phase space of water drop impact onto polymeric microstructures. Different colors indicate different roughness, \Rf~$ = 1.9~\text{(green)},~2.6~\text{(magenta)},~3.0~\text{(black)},~\text{and}~5.0~\text{(cyan)}$, determined by the geometric arrangements of the micropillars of $h = 10 \unit{\mu m}$. Different symbols present various dynamical impact behaviors:  jetting ($\ast$), complete rebound ($\circ$), partial rebound ($\triangle$), sticky ball ($\Box$), formation of a few satellite droplets while the main impact reveals a partial rebound ($\rhd$) or a sticky wetting ball ($\lhd$). }
\end{center}
\end{figure}

Fig.~\ref{PDMS_We120} shows a partial rebound for an impacting drop, with similar control parameters $P_{air} = 101.3~\unit{kPa}$, \We~$= 120$ and \Rey~$= 2882$, but now for a polymeric microstructured surface consisting of round pillars of $a=1.27~\mu m$, $w=5~\mu m$ and $h=10~\mu m$. 
As observed from the snapshots, the rim is not destabilized in the same way as for the unstructured CNFJs (compare Fig.~\ref{splashing_figs} and~\ref{PDMS_We120} at $t \approx 1$ \unit{ms}). Instead of perimetric fingers, wavy bumps develop and eventually merge together at the center to form an elongating water fountain. The pinning of the water wets the microstructured surface and breaks the fountain while surface tension is in action, resulting in a few breaking-up droplets. 
Consistent with the results in Ref.~\cite{DQuere_bouncing_transition_EPL2006}, our experiment at \We~$=147$, \Rey~$=3524$ and $P_{air} = 101.3~\unit{kPa}$ shows that a few ($\approx 5$) satellite droplets develop during the contraction of the water sheet upon a micropatterned substrate of $a=5~\unit{\mu m}$, $w=5~\unit{\mu m}$ and $h=20~\unit{\mu m}$ with a roughness \Rf $= 2.8$. The comparisons between these different experiments reveal that roughness has a profound effect on splashing impacts. In addition, the multiscale, uncontrolled surface roughness of CNFs display pronounced formations of satellite droplets, comparable with the impacts upon ordered micropatterned substrates under alike parameters. This finding suggests that the smaller length scale of carbon fibers may enhance the instability of perimetric wavy fingers and thus produce stronger fragmentations. 

Fig.~\ref{PDMS_phase_diagram} displays the phase space of water impact dynamics onto polymeric microstructures of $\theta_{SCA}=(147 \pm 3)^{\circ}$ explored at different air pressure $P_{air}$.  Different colors depict different surface roughness \Rf~controlled by the geometric arrangement of micropillars of $h = 10 \mu m$. Similar impact phenomena to those upon CNFs are observed, including jetting, complete rebound, and partial rebound for small \We~$\apprle 10$.  For $70 \apprle$ \We $\apprle 110$ partial rebound occurs with a high deformability stretching the water droplet into a elongating column. 
% [changes]
As discussed in Ref.~\cite{DQuere_bouncing_transition_EPL2006}, a threshold of impact velocity marking a transition from complete bouncing to wetting or partially pinned drop is identified as a function of the size of the superhydrophobic microtextures.  Interestingly, for the small roughness \Rf~$= 2$, we observe coexistence of different dynamical processes in some range of impact velocities.
At large \We~(between $200$ and $300$) a few satellite droplets form while the main impact can partially rebound ($\rhd$) or behave like a sticky wetting ball (shown by $\lhd$).

% ===================
\section{Conclusions}
% ===================

In comparison with the drop impacts performed upon microstructured hydrophobic surfaces, for \We~$\apprle 10$ and under the atmospheric pressure impacts upon CNFs of multiscale roughness reveal similar dynamical behaviors including Cassie-wetting Fakir droplet, complete rebound, partial rebound, jetting, and trapping bubbles. This comparison implies that impact evolutions are insensitive to the details of the roughness of superhydropohobic surfaces at small \We-number.   
Rather than a bouncing-pinning transition, driven by a threshold of impact velocity, as described in Ref~\cite{DQuere_bouncing_transition_EPL2006} for regular microstructures, we noticed the coexistence of complete rebound and sticky wetting droplet for CNFs at $1.1 \apprle$ \We~$\apprle 1.9$.  This observation indicates that the details of roughness can influence the transitional boundary in the phase space of impact events. 

Partial rebounds occur in the \We-range between $10$ and $130$ for water drops hitting upon micropatterned substrates with $h = 20~\unit{\mu m}$ and a roughness ranging from $2$ to $10$. % for PDMS 2920.
Air pressure and the surface roughness have hardly any effect in this partially rebouncing regime. At higher \We~$\apprge 120$, profound splashing impacts forming several satellite droplets take place for superhydrophobic CNFs. This splashing mechanism is not suppressed by the decrease in air pressure, which was found to debilitate the corona splashing on smooth wetting surfaces~\cite{xu:splashing_PRL2005,xu_PRE2007_2}. 
Beyond \We~$\apprge 140$, in contrast to the violent splashing events on the rough CNFJs, on imcrostruture substrates the fragmentation is much less pronounced, revealing the importance of sub-micron roughness for splashing.

\section{Acknowledgments} The authors thank Alisia M. Peters and Hrudya Nair for their helps on the preparations of the polymeric microstructures.  We also gratefully acknowledge the Membrane Science and Technology Group at the University of Twente for the usage of the contact angle measurement device.

\bibliography{impact_upon_CNFs_submitted_version_Jan26_2009}

\end{document}